\documentstyle[12pt]{article}
\makeatletter
\newcommand{\singlespacing}{\let\CS=\@currsize\renewcommand{\baselinestretch}
{1.0}\tiny\CS}
\newcommand{\doublespacing}{\let\CS=\@currsize\renewcommand{\baselinestretch}
{1.5}\tiny\CS}

\begin{document}

\title{ Two theorems of Jhon Bell and Communication Complexity}
\author{Guruprasad Kar\\
Physics and Applied Mathematics Unit,\\ Indian Statistical
Institute,\\ 203 B. T. Road, Kolkata 700108,\\
India\\
{\bf email : gkar@isical.ac.in}}

\date{}

\maketitle

\vspace{0.5cm}

\begin{center}
{\bf Abstract}
\end{center}

John Bell taught us that quantum mechanics can not be reproduced
by non-contextual and local Hidden variable theory. The
impossibility of replacing quantum mechanics by non-contextual
Hidden Variable Theory can be turned to a impossible coloring
pseudo-telepathy game to be played by two distant players. The
game can not be won without communication in the classical world.
But if the players share entangled state (quantum correlation) the
game can be won deterministically using no communication. This
again shows that though quantum correlation can not be used for
communication, two parties can not simulate quantum correlation
without classical communication. The motivation of the article is
to present the earlier works on Hidden Variable Theory and
recently developed pseudo-telepathy problem in a simpler way,
which may be helpful for the  beginners in this area.

\section*{Introduction}

Quantum mechanics is a mathematical theory to describe the physical
world and at the same time it is a probabilistic theory. But this is
not surprising. What is surprising, if we take Copenhagen
interpretation for granted, is that this probability is not the
probability of some dynamical variable having a particular value in
some state, but that represents probability of finding a particular
value if that dynamical variable is measured. This interpretation
generated numerous debates and following Einstein[1] many people
used to think that in future quantum mechanics will be replaced by
some more fundamental theory in which description of state will be
complete in the sense that every dynamical variables will have well
defined values and  measurement merely reveals those pre-existing
values.

If the question is put forward in this way; Can quantum mechanics be
replaced by some complete theory in the above sense? At least the
uncertainty principle and complementarity principle do not resolve
this question in the negative. Uncertainty relation tells that if
one prepares an ensemble corresponding to a quantum state, then all
members of that ensemble can not have well defined values for the
observables  (like position and momentum) involved in the
uncertainty relation but it is mute regarding the question whether
individual member can have well defined values for observables [2].
Again complementarity principle prohibits joint measurement of
certain observables (like position and momentum, polarization
measurement in different directions)without addressing the question
whether the system has well defined values of those observables.

For a long time, there was a general belief among quantum physicists
that quantum mechanics can not be replaced by some complete theory,
also popularly known as Hidden Variable Theory(HVT), due to Von
Neumann's impossibility proof (who imposed an unwarranted constraint
on HVT). In sixties we got two theorems due to J.S.Bell and Kochen
and Specker [3,4,5]. This two theorems showed that quantum mechanics
can not be replaced by some classes of HVT, namely local and
non-contextual HVT.

Here we shall present the impossibility of replacing quantum
mechanics by some non-contextual Hidden Variable Theory by an
example in four dimensional Hilbert space [6]. Using this example
we shall construct a game to be played by two distant players who
are deprived of having any type communication. The impossibility
proof itself will provide the reason why in classical world the
players can not win the game with certainty. But we shall see that
they can win it by some protocol (not involving classical
communication)if they share some quantum correlation
beforehand[7]. This, in a fundamental way, proves the celebrated
Bell's theorem that no local Hidden Variable Theory can replace
quantum mechanics[5].

\section*{Description of Hilbert Space Quantum Mechanics}

\begin{enumerate}
\item Every system is associated with a Hilbert space.
\item States are associated with vectors in the Hilbert space.
\item Observables are associated with self adjoint operator acting
on the corresponding Hilbert space.
\end{enumerate}

Consider a self adjoint operator $A$ acting on the finite
dimensional($n$, say) Hilbert space H .Operator $A$ will have n
eigenvalues (different in general) $a_1, a_2, ....a_n$ and let the
corresponding eigenfunctions are $|\psi_1\rangle,
|\psi_2\rangle....|\psi_n\rangle$. Equivalently $A$ can be written
as (also known as spectral representation ) $$A = \sum
{a_i|\psi_i\rangle\langle\psi_i|}$$ where
$|\psi_i\rangle\langle\psi_i|$ is a one dimensional projection
operator. In general a projection operator $P$(say) has two
eigenvalues 1 or 0 as $P$ is idempotent($P^2 = P$).\\

\section*{Measurement}

\begin{enumerate}

\item Possible measurement result is one of the eigenvalue of the
observable  measured on the system.

\item If the initial state of the system is $|\psi\rangle$, and the
observable that is measured is $A$, then the probability that the
measurement result will be $a_k$ is given by
$$Prob_{|\psi\rangle}(a_k) = Tr[|\psi\rangle\langle\psi||\psi_k\rangle\langle\psi_k|] =
|\langle\psi|\psi_k\rangle|^2$$ As the eigenvalue does not appear in
the probability expression this can be equivalently expressed in the
following way; If measurement is performed in the basis
\{$|\psi_i\rangle$\}, which means to decide which of the projector
of the complete set
\{$|\psi_i\rangle\langle\psi_i|~;~\sum|\psi_i\rangle\langle\psi_i| =
I$\} is true, then the probability that the value of the projector
$|\psi_k\rangle\langle\psi_k|$ will be 1(true) is  given by the same
quantity as
$$Prob_{|\psi\rangle}(|\psi_k\rangle\langle\psi_k| = 1) = |\langle\psi|\psi_k\rangle|^2$$
In general, if one performs a measurement to decide whether the
projector $P$ is true(1) or false(0), then the probability that
measurement result of $P$ will be true, is given by
$Tr[|\psi\rangle\langle\psi|P] = \langle\psi|P|\psi\rangle$

\item After the measurement the state will collapse on the
corresponding eigen-state i.e. if the result is $a_k$ (or
equivalently $|\psi_k\rangle\langle\psi_k| = 1$) then final state
will be $|\psi_k\rangle$. In general, if the measurement is to
decide whether a projector $P$ is true or false, and if the result
corresponds to the truth of $P$, then the final states will be
$\frac{P|\psi\rangle}{|P|\psi\rangle|}$
\end{enumerate}

\section*{Unitary dynamics}

The future development of a state is given by the unitary dynamics
where the unitary operator $U$ ($UU^{\dag} = U^{\dag}U = I$)is
determined by the Hamiltonian acting on the system.The dynamical
equation is given by the celebrated Schrodinger equation
 $$|\psi(t)\rangle = U(t,t_0)|\psi(t_0)\rangle$$
Unitary operator preserves the scalar product between any two
vectors i.e. $<\psi_1|\psi_2> = <\psi_1|UU^{\dag}|\psi_2>$. If two
vectors $|\psi_1>$ and $|\psi_2>$ are orthogonal then $U|\psi_1>$
and $U|\psi_2>$ are also orthogonal.

\section*{Is there underlying deterministic theory ?}

If we prepare two systems in the same state  $|\psi\rangle =
c_1|\psi_1\rangle + c_2|\psi_2\rangle$ ($c_1,c_2 \ne 0$)  and on
both the observable $A$ is measured, the result ,in general, will
not be identical as probability for the results  $a_1$ and $a_2$ are
$|\langle\psi|\psi_1\rangle|^2$ ($ = |c_1|^2$) and
$|\langle\psi|\psi_2\rangle|^2$ ($ = |c_2|^2$) respectively, both
being non-zero. This indeterminism is fundamental, because according
to quantum mechanics, the initial quantum states were truly
identical. Having a underlying deterministic theory would imply that
actually two states were not identical, they differ in some
additional parameters (popularly known as hidden variables) which
remain unknown to us. If description of state is to be completed,
these parameter has to be taken into account and then the theory
will behave in a deterministic way.\\

\section*{Programme of Hidden Variable Theory}

 There exists a complete theory where states
assign values to all observables (equivalently yes/no to all
projectors) and quantum state is some kind of statistical mixture of
these finer states. Schematically, in quantum mechanics;
$$Prob_{|\psi\rangle}(P = 1) = \langle\psi|P|\psi\rangle \ne 0~ or~1$$ in general.\\
Now if the state $|\psi\rangle$ is completed with the additional
variable say $\lambda$, then the completed state $|\psi,
\lambda\rangle$ (not a vector in the Hilbert space) will be able to
assign values to all projectors when value of $\lambda$ is
specified. It is the incapacity of quantum theory that it treats two
different state $|\psi, \lambda_1\rangle$ and $|\psi,
\lambda_2\rangle$ ($\lambda_1 \ne \lambda_2$) as same state. So we
want a complete theory(HVT) where if the value of $\lambda$ is known
then probability of every proposition for that completed state
becomes 0 or 1, and hence we can talk in terms of proposition being
true or false (projector having value 0 or 1) instead of
probability. Hence

$$Prob_{|\psi, \lambda\rangle}(P = 1) =   V_{|\psi, \lambda>}(P) = 0 ~or~ 1$$
 In this picture quantum state $|\psi\rangle$ is simply a statistical
mixture of completed states $|\psi, \lambda\rangle$ and quantum
probability arises due to ignorance on the value of $\lambda$. So if
the distribution of $\lambda$ is given by $\rho(\lambda)$ with
$\int\rho(\lambda)d\lambda = 1$, then the quantum probability will
arise due to ignorance on $\lambda$, which can be expressed
mathematically as
$$\langle\psi|P|\psi\rangle = \int\rho(\lambda)V_{|\psi, \lambda\rangle}(P)d\lambda$$
Is it possible to have a HVT even in principle?

Bell showed that if the dimension of the Hilbert space is two, then
one can construct a HVT which reproduces QM so far as standard
measurement represented by projector is concerned[3].

For two dimensional Hilbert space, there is a one to one association
between states and points on the surface on a unit sphere. For any
state $|\psi\rangle$, there is a unit vector $n$, such that the
projector on $|\psi\rangle$ can be written as,\\
$$|\psi\rangle\langle\psi|
= \frac{1}{2}~[I +n.\sigma]$$ where $n$ is the unit vector,
$\sigma$'s are  $2\times 2$ Pauli matrices  and $\sigma.n$ = $
n_x\sigma_x + n_y\sigma_y + n_z\sigma_z$ and $I$ is an unit operator
on 2 dimensional Hilbert space.
 Now consider a projector $P$ where   $$P = \frac{1}{2}~[I +m.\sigma]$$
Then
$$Prob_{|\psi\rangle}(P = 1) =\frac{1}{4}~Tr[(I +n.\sigma)
(I +m.\sigma)] = \frac{1}{2}~[1 +n.m]$$
which can take values from 0
to 1 depending on the scalar product of $m$ and $n$.\\

\section*{Construction of HVT in two dimension}

 Let us now consider completed (HVT) state
$|\psi,\lambda\rangle$ or $|n,\lambda\rangle$, where the additional
variable $\lambda$ varies from $-\frac{1}{2}$ to
$\frac{1}{2}$, and distribution of $\lambda$ is uniform.\\
Now our completed state has to assign values 0 or 1 to all
projectors. Let the state assign values to the projectors in the
following way,
$$V_{|n,\lambda\rangle}[P] = \frac{1}{2}~[1 + Sign
(\lambda + \frac{1}{2}~|n.m|~ Sign(n.m))]$$\\
One can easily check that it reproduces the quantum probability if
one integrates over $\lambda$ [8]. Let us now calculate quantum
probability for the case for which $n.m$ = -ve
$$ \begin{array}{lcl}
Prob_{|\psi\rangle}(P = 1)&=&\displaystyle
\int_{-\frac{1}{2}}^{\frac{1}{2}}{\rho(\lambda)V_{|n,\lambda\rangle}(P)d\lambda}\\
&=&\displaystyle{ \int_{-\frac{1}{2}}^{\frac{1}{2}}{\frac{1}{2}~[1
+ Sign(\lambda + \frac{1}{2}~|n.m|~ Sign(n.m))]d\lambda}}\\
&=&\displaystyle{
\frac{1}{2}\left[\int_{-\frac{1}{2}}^{\frac{1}{2}|n.m|}{(1 -
1)~d\lambda} + \int_{\frac{1}{2}|n.m|}^{\frac{1}{2}}{(1 +
1)~d\lambda}\right]}\\
&=&\displaystyle{ \frac{1}{2}[1 - |m.n|]}\\
&=&\displaystyle{\frac{1}{2}[1 + (m.n)]}\\ \end{array}$$

Similarly one can check it, for the case when $n.m$ = +ve

\section* {Is HVT possible in higher dimension ?}

After successful HVT model in two dimension, the natural question
arises whether the construction is  possible in higher dimension. In
higher dimension there is further question of context which did not
arise for two dimensional Hilbert space. For simplicity, consider a
three dimensional Hilbert space where $\{|\psi_i\rangle\}_{i =
1,2,3}$ is a orthogonal basis. Then HVT assigns 1(truth) or 0(false)
value to
all the projector with the following restriction;\\
\begin{enumerate}
\item $V(|\psi_i\rangle\langle\psi_i|) = 0 ~or~ 1$\\
\item As the projectors are mutually exclusive, at a time one of
them can be true(1) and the rest will be false(0) which means
$\sum V(|\psi_i\rangle\langle\psi_i| = 1$\\
\end{enumerate}
Now consider another orthogonal basis $\{|\psi_1\rangle,
|\phi_2\rangle, |\phi_3\rangle\}$ where $|\phi_2\rangle =
\frac{1}{\sqrt2}(|\psi_2\rangle + |\psi_3\rangle)$ and ,
$|\phi_3\rangle = \frac{1}{\sqrt2}(|\psi_2\rangle -
|\psi_3\rangle)$. Obviously measurement in this basis is different
from measurement in the basis considered earlier. But the projector
$|\psi_i\rangle\langle\psi_i|$ is common. Does the HVT assign same
value (0 or 1) to this common projector ignoring the different
measurement context. If the answer is yes, then the HVT is called
non-contextual(One should note that this question did not arise in
two dimension as there can not be two different orthogonal basis
having one vector in common).

Then it has been shown that this kind of Hidden Variable Theory
namely non-contextual HVT  is impossible if the dimension of the
Hilbert space is more than or
equal to three [3,4].\\
We shall present a proof of this impossibility for four dimensional
Hilbert space[6]. We consider the following 18 (unnormalized)
vectors appearing in 9 sets of orthogonal basis and assign value (0
or 1)to the projectors on the corresponding vectors under the
restriction (a),(b) along with the non-contextual assumption. For
short, we have used the symbol $V(|\psi>)$ to mean the value of the
projector $V(|\psi\rangle\langle\psi|)$


$$ V(0 0 0 1)+ V(0 0 1 0)+ V(1 1 0 0)+ V(1 -1 0 0) = 1$$

$$ V(0 0 0 1)+ V(0 1 0 0)+ V(1 0 1 0)+ V(1 0 -1 0) = 1$$

$$ V(1 -1 1 -1)+ V(1 -1 -1 1)+ V(1 1 0 0)+ V(0 0 1 1) = 1$$

$$ V(1 -1 1 -1)+ V(1 1 1 1)+ V(1 0 -1 0)+ V(0 1 0 -1) = 1$$

$$ V(0 0 1 0)+ V(0 1 0 0)+ V(1 0 0 1)+ V(1 0 0 -1) = 1$$

$$ V(1 -1 -1 1)+ V(1 1 1 1)+ V(1 0 0 -1)+ V(0 1 -1 0) = 1$$

$$ V(1 1 -1 1)+ V(1 1 1 -1)+ V(1 -1 0 0)+ V(0 0 1 1) = 1$$

$$ V(1 1 -1 1)+ V(-1 1 1 1)+ V(1 0 1 0)+ V(0 1 0 -1) = 1$$

$$ V(1 1 1 -1)+ V(-1 1 1 1)+ V(1 0 0 1)+ V(0 1 -1 0) = 1$$


Add these nine equations. The left hand side will be even as every
vector has appeared twice, while the right hand side is obviously
odd.\\

\section*{Entangled state}

Let us now consider two systems A and B, both associated with
d-dimensional Hilbert space and let
$\{|i\rangle_{A(B)}\}_{i=1..d}$ is a orthogonal basis for A(B).
Consider the state
$$|\phi_{AB}\rangle = \frac{1}{\sqrt{d}}~\sum_{i=1}^d |i\rangle_A \otimes |i\rangle_B$$ One can
see that this state $|\phi_{AB}\rangle$ can not be written in the
form of product state $|\chi\rangle_A \otimes |\gamma\rangle_B$.

A state of joint system that can not be written as product of
subsystem states, is called entangled state.\\
Consider a Unitary  $d\times d$ matrix $U$ acting on d-dimensional
Hilbert space. Then $\{U|i\rangle_{A(B)}\}_{i=1..d}$ is a new basis.
Again $U^\ast$ (the new operator formed by replacing each element by
its complex conjugate)is also a unitary operator which also generate
another orthogonal basis if
acted on a orthogonal basis.\\
Then the state $|\phi_{AB}\rangle$ has the following  interesting
property;
$$U_A \otimes U_B^\ast~|\phi_{AB}\rangle = |\phi_{AB}\rangle$$
In case $U = U^\ast$, we get
$$|\phi_{AB}\rangle = \frac{1}{\sqrt{d}}\sum_{i=1}^d |i\rangle_A \otimes |i\rangle_B =
\frac{1}{\sqrt{d}}\sum_{i=1}^d U|i\rangle_A \otimes U|i\rangle_B$$
So $|\phi_{AB}\rangle$ can be written in terms of any orthogonal
basis if it is related to $\{|i\rangle\}$ by a unitary operator $U$
such
that $U = U^\ast$.\\

\section*{Bi-partite pseudo-telepathy}

Consider the previous 9 sets of orthogonal basis in four dimension
and denote them by $S^1, S^2.....S^9$, where $S^J$ contains the
following four orthogonal vectors $|u_1\rangle^J$,
$|u_2\rangle^J$, $|u_3\rangle^J$, $|u_4\rangle^J$ where,
$|u_1\rangle^1 = |u_1\rangle^2 = (0 0 0 1)$, $|u_2\rangle^1 =
|u_1\rangle^5 =
(0 0 1 0)$,etc.\\
Two players, say Alice and Bob, are far away and (1) Alice is
given any one ($S^k$, say) of the nine sets randomly, and (2) Bob
is given a vector ($|u_m\rangle^k$, say) randomly from the set
given to Alice.

\section*{ Winning condition}

Alice has to assign value (0 or 1) to her four vectors and Bob
also
has to assign value to his single vector in such a way that \\
\begin{enumerate}
\item Exactly one of Alice's vector should receive the value 1.\\
\item Alice and Bob have to assign same value to their single
common vector\\
\end{enumerate}
with the condition that they will not be allowed to have any
classical communication after the game starts. They will win the
game if in every repetition of the game they win it with certainty.\\

\section*{Players in the classical world}

Any classical deterministic winning strategy would mean precisely
assigning values to all the 18 vectors in a non-contextual way which
is already been forbidden by the Bell and Kochen-Specker theorem. No
previously shared correlation can help them in this regard.

If one tries to satisfy all the equations by assigning
non-contextual values to the maximum possible no. of vectors, then
one would see that 17 vectors can be assigned non-contextual values
and value assignment for one vector has to be contextual i.e. one
vector out of 18 has to take value 1 when it occurs in one basis and
0 when it occurs in another basis.

Let us consider a contextual solution  where the vector $(0 0 0 1)$
takes value 1 when it occurs in $S^1$ and 0 when in $S^2$. Now one
can have simplest strategy like this. Alice sends bit 0 to Bob when
he is given the basis set $S^1$ and bit 1 when it is not $S^1$, so
that they can win the game by assigning value to the vector $(0 0 0
 1)$in a contextual way. So 1 cbit is more than sufficient to win
the game in the classical world.

\section*{Quantum strategy}

Let Alice and Bob share the following entangled state
$$ |\psi\rangle_{AB} = \frac{1}{2}[|u_1\rangle_A^1|u_1\rangle_B^1 + |u_2\rangle_A^1|u_2\rangle_B^1 +
|u_3\rangle_A^1|u_3\rangle_B^1 + |u_4\rangle_A^1|u_4\rangle_B^1]$$
But all the 18 vectors involved here are real and hence all the
other 8 orthogonal basis sets $S^2, S^3,...., S^9$  are related to
this one $\{|u_i\rangle^1\}$ by unitary operator whose matrix
elements are real. So the state can be written in any of the basis
keeping the form of the state intact (overall phase is omitted) i.e.
$$|\psi\rangle_{AB} = \frac{1}{2}\sum_{i=1}^4[|u_i\rangle_A^k|u_i\rangle_B^k]$$
for all $k$.

{\bf Alice}: She measures on her system in the basis of the set, she
has been given. On whatever state she collapses due to measurement,
she assigns value 1 to that vector and  assigns 0 to the rest three.

{\bf Bob}: He chooses any basis set containing the vector given to
him and measure in that basis. If he collapses on the vector given
to him, he assigns 1 and 0 otherwise.

With this strategy they can win the game deterministically.

Consider one example where Alice has been given the set $S^m$ and
Bob has been given the vector $|u_p\rangle^m$. Now the shared state
can be written  in the basis $S^m$ as
$$|\psi\rangle_{AB} = \frac{1}{2}\sum_{i=1}^4(|u_i\rangle_A^m|u_i\rangle_B^m)$$
Now according to the protocol described above,  Alice will measure
in the basis ($|u_1\rangle^m, |u_2\rangle^m, |u_3\rangle^m,
|u_4\rangle^m$). Let due to the measurement Alice's system collapses
on the vector $|u_q\rangle^m$. Because of the  correlation Bob's
system will also collapses on $|u_q\rangle^m$. Then due to Bob's
measurement, the probability that his system will collapse on the
vector $|u_p\rangle^m$, given to him, is given by
$$Prob_{Bob}(|u_p\rangle^m) = |\langle u_p|u_q\rangle|^2 = \delta_{pq}$$
(Actually there is no temporal order between Alice's and Bob's
measurement in the protocol. The result will remain same whoever
performs the measurement earlier.) This clearly shows that Bob will
assign value 1 to the vector given to him only when he collapses on
that vector and it can happen if and only if Alice also collapses on
that vector. So the protocol will work
in every cases without error.\\

\section*{What is the teaching?}

\begin{enumerate}
\item One should note that, though there there is a successful
protocol with entanglement, Bob can not know which set his vector
belongs to (recall that every vector has appeared in two different
basis sets). If he could know it that would  imply signalling or
real telepathy. Entanglement can offer advantage for only a class
of problems of this type, solving which itself does not imply
signalling (transfer of real information from one party to another
instantaneously) or violation of causality.\\
\item The game can not be won deterministically in the classical world
without classical communication. So it shows that classical
correlation (shared randomness) can not generate the effect of
quantum correlation if no classical communication is allowed.This
is the essence of most celebrated Bell's theorem, which says that
no local Hidden Variable theory can reproduce all the correlations
of quantum mechanics.\\
\item Communication complexity is an area that aims at quantifying the
amount of communication necessary to solve distributed
computational problem. Quantum communication complexity uses the
power of entanglement to reduce the amount of communication that
would be classically required. Here we discuss pseudo-telepathy
problem to exhibit  the potential power of entanglement to
penetrate in the area of communication complexity of course under
the condition of no violation of causality. This kind of example
clearly indicates that for cleverly chosen distributed
computational problem, entanglement can be used to reduce the
amount of bits of classical communication needed to solve it in
the classical world.
\end{enumerate}

\section*{Acknowledgement}

I thank Anirban Roy and Debasis Sarkar for helpful discussion. I am
grateful to Samir Kunkri who pointed out  that contextual value
assignment to a single vector out of 18 satisfies all the 9
equations used in the impossibility proof.\\

\section*{References}

\begin{enumerate}
\item A.Einstein, B.Podolsky and N. Rosen {\it Phys.
Rev.} {\bf 85}, 777 (1935).\\

\item N. D. Mermin {\it Rev.Mod. Phys. } {\bf 65}, 803 (1993).\\

\item J.S. Bell {\it Rev.Mod. Phys. } {\bf 38}, 447 (1966).\\

\item S. Kochen and E.P. Specker {\it J. Math. Mech. } {\bf 17}, 59
(1967).\\

\item J.S. Bell {\it Physics 1 } 195-200 (1964).\\

\item A. Cabello, J.M. Estebaranz and G. Garcia-Alcaine  {\it Phys. Lett. A } {\bf 212 }, 183 (1996).\\

\item G. Brassard, A. Broadbent and A. Tapp,  quant/ph-0407221.\\

\item `The logic of quantum mechanics' by E.G.Beltrametti and
G.Cassinelli, Addison-Wesley Pub.Co., 1981.
\end{enumerate}

\end{document}